\documentstyle[12pt,fleqn]{article}
\newcommand{\rf}[1]{\raisebox{1ex}{\cite{#1}}}

\newcommand{\fg}[1]{\item {\label{#1}}}

\begin{document}

\vskip 40pt

\centerline {\bf Symbolic dynamics III}   
\centerline {\bf Bifurcations in billiards and smooth potentials}
\vskip 1.5cm                                                               
  
\centerline { 
Kai T. Hansen
           }
\centerline {\em Niels Bohr Institute
\footnote{
{\ddag}
{\small Permanent address: Phys Dep., University of Oslo, Box 1048, Blindern,
   N-0316 Oslo}
} 
}
\centerline {\em Blegdamsvej 17, DK-2100 Copenhagen \O}  
\centerline {{\em e-mail:}  khansen@nbivax.nbi.dk}

\vskip 25pt
{\centerline{\bf ABSTRACT}}
\noindent{
The singular bifurcations in a dispersive billiard are discussed in
terms of symbolic dynamics and is compared to an example of a
bifurcation tree in a smooth potential.  Possible generalizations to
other smooth potentials are discussed.
   }
 
\vfill\eject

\section{Introduction} 

The aim of this article is to describe the bifurcation structure of some
families of 2-dimensional Hamiltonian systems.  Bifurcations in
one dimensional unimodal maps are well understood, and the Metropolis,
Stein, Stein (MSS)\rf{mss} theory uses symbolic dynamics to establish
that the order of  bifurcations is universal for all unimodal maps.  
However, there is no general theory of bifurcations in Hamiltonian
systems analogous to the MSS theory for unimodal maps.  The method
presented here predicts the bifurcations of a  periodic orbit, including
all periodic orbits emanating from a bifurcation, and the non-periodic
(chaotic) orbits born at the accumulation points  of bifurcations.  We
refer to these orbits as the members of one bifurcation family, and we
show that the members of a family are well  defined and do not change
with small changes of the Hamiltonian.  However, the ordering in the parameter
space of the bifurcations within a family may be different for
different smooth potentials, and the ordering of different families
in the parameter space is  not known a priori.

We are not able to define rigorously the class of smooth Hamiltonians to
which the method described here applies,  but the Hamiltonian must have some
similarities with a billiard system.  This includes physical interesting
examples such as the classical collinear $CO_2$ molecule\rf{co2}, and
possibly the classical collinear {\em He} atom\rf{wintgen} and the 
St\o rmer problem\rf{jung}. 
The smooth potential we use as an example is (for reference see
\rf{DR_prl,DR})
\begin{equation}
  H = \frac{1}{2} \left( p_x^2 + p_y^2 + (x^2y^2)^{{1}/{a}} \right) \label{e_H}
\end{equation}
where variation of the parameter $a$ gives rise to a family of
Hamiltonians.  In the limit $a\rightarrow 0$ the system reduces to the
hyperbola billiard and all orbits can be assigned a
well defined symbolic dynamics\rf{hyperbola,cvi_eck_sym}. 
As $a$ increases, orbits disappear through inverse bifurcations.  The
bifurcations are period doublings, symmetry breaking, tangent
bifurcations and bifurcations of higher order, and they always imply
stable orbits; hyperbolic orbits  become elliptic orbits for some
parameter interval and then disappear\rf{DR}, as shown in the examples
below.

A billiard system, e.g. the 3-disk billiard discussed below, with a
parameter $r$, also has a bifurcation scenario.  For a parameter value
$r_c$, an infinite number of orbits are pruned, but in the billiards 
considered here, no orbits become stable.

We begin by reviewing the bifurcations in one dimensional unimodal maps.
We then describe the simplest example in a 3-disk billiard, and  discuss
in detail a bifurcation sequence in the 4-disk billiard, and compare it 
with the bifurcations found by  Dahlqvist and Russberg for the
Hamiltonian (\ref{e_H}).


\section{Unimodal maps}

Here we motivate our approach by showing that bifurcations in a 
one-dimensional logistic map are analogous to bifurcations in a smooth
potential,  and that bifurcations in the tent map are analogous to
bifurcations in a billiard. The logistic map is defined as  
\[ x_{t+1} = ax_t(1-x_t) \] 
\noindent and the symbolic dynamics we use is symbol 1 if $x>1/2$ and
symbol 0 if  $x<1/2$.  The MSS bifurcation scenario starts out as
follows: at $a=1$ the fixed point $\overline{1}$ becomes stable, at 
$a=3$ the fixed point $\overline{1}$ becomes unstable  while the period
2 orbit $\overline{10}$ is born as a stable orbit, at $a=3.4494\ldots$
the orbit $\overline{10}$ becomes unstable and the period 4 orbit
$\overline{1011}$ is born as a stable orbit, and so on.   The MSS theory 
predicts period doubling of an orbit  into a periodic orbit of twice the
length.  The symbolic dynamics description of the new orbit is  obtained
by writing the original symbol sequence twice and then changing the last
symbol from 0 to 1 or from 1 to 0; orbit $\overline{1011}$ period
doubles into $\overline{10111010}$, etc.  At the accumulation point
$a=3.5714\ldots$ all such orbits exist and are unstable.  We call this
collection of orbits the period doubling family of the fixed point
$\overline{1}$.   In the same way one finds the period doubling family
of the period 3 orbit $\overline{100}$, and other orbits born by tangent
bifurcations. To be more precise, we can identify symbol 0 with $x<1/2$
and symbol 1 with $x>1/2$ for stable orbits in this family, only when
the orbit has passed through  the superstable point $x=1/2$.  Between
the bifurcation point, where the orbit is born, and the superstable
point, the naive symbolic description of the stable orbit is the same as
for a different unstable orbit.  This unstable orbit is either an orbit
of the same length created in a tangent bifurcation, or the orbit which
became unstable at a period doubling bifurcation.  

In the tent map
\[   x_{t+1} = \left\{ \begin{array}{ll}
                        r x_t & x \leq 1/2 \\
                        r(1-x_t) & x > 1/2
                     \end{array}
             \right.  \]
\noindent this bifurcation scenario is slightly altered. The fixed point
$\overline{1}$ exists 
only when $r>1$.  Since the slope at this fixed point is $-r<-1$, the
fixed point is unstable  and the period 2 orbit  $\overline{10}$ exists
when $r>1$.  This  orbit has stability $-r^2<-1$, and is also unstable. 
By induction all the orbits in the period doubling family of
$\overline{1}$ exist and 
are unstable for $r>1$ and consequently $r_c=1$ is the critical
parameter value where all orbits in this family bifurcate simultaneously. 
For $r=r_c$ all points in all orbits have $x$-values equal $1/2$ and when
$r$ increases from $r_c$ the points in the orbits spread out on the
$x$-axis. The critical value for the period 3 family $\overline{100}$
is $r_c = (1+\sqrt{5})/2=1.6180\ldots$, with all orbits in the period
doubling family born unstable, and the same happens for all orbits born
in tangent bifurcations.  
There may also be other orbits born at these critical parameter values.

MSS universality is a consequence of the fact that the parameter space
is one-dimensional for a unimodal map, and there is only one possible
ordering of bifurcations for any monotone variation of the parameter. 
An $n$-modal one-dimensional map has more complicated bifurcations in an
$n$-dimensional parameter space. For higher dimensional maps, the
situation
is totally different; the parameter space is in principle always
infinite dimensional.

\section{Bifurcations in disk billiards and potentials} 

The billiards that we use here as our standard examples of dispersive
billiards consist of a point  particle scattering elastically between
$N$ circular disks in the plane.  The basic symbolic description is
obtained by enumerating the disks, and letting a bounce off disk number
$s$ at the integer time $t$ be denoted by the symbol $s_t$.  Other
symbols can be made from $s_t\in\{1,\ldots ,N\}$, the number of symbols
can be reduced\rf{DR,cvi_eck_sym}, and a well ordered alphabet can be
constructed\rf{hansen_1,hansen_2}, but for simplicity we
shall use only the basic symbols here. 

The simplest chaotic billiard consists of 3-disks of radius 1, with the
distances between  the centers of the disks given by parameters
$r_{12}$, $r_{13}$ and $r_{23}$.  If the disks are sufficiently
separated\rf{hansen_2}, all orbits  that can be described by any
infinite symbol sequence $\ldots s_{t-1}s_t s_{t+1}\ldots$ exist, except
orbits with two successive bounces off the same disk.  As shown in
ref.\rf{hansen_1,hansen_2}, a bifurcation in the billiard takes place
when a free path between two  bounces is tangent to a disk, or when the
bounce of the particle has an incoming and outgoing direction parallel
to the edge of the disk.  If the parameter $r$ changes further the orbit
is not admissible in the system; we call such orbit pruned.  It is clear
that at the critical
bifurcation parameter value $r_c$ the orbit that is tangent to the disk
is indistinguishable in configuration space from the orbit that bounces
parallel to the edge of the disk. From this fact we can establish the
set of orbits that bifurcate at $r_c$ and at the same point in phase
space.  Analogously to the one-dimensional tent map we call this set of
orbits a bifurcation family.

Consider as an example the periodic orbit $\overline{123132}$ drawn in
figure~\ref{f_123132}~a) for $r_{12}=r_{13}=r_{23}=2.5$.  To obtain
simple figures we have chosen the  distance between the centers of the
disks 2 and 3 to be the parameter $r=r_{23}$ while keeping $r_{12}=r_{13}=2.5$. 
This orbit bifurcates at $r_c=4.1044\ldots$, as shown in
figure~\ref{f_123132}~b) where the line from disk 2 to disk 3 is tangent
to disk 1, and the orbit is not admissible for $r>r_c$.  
Figure~\ref{f_123132}~c) shows the periodic orbit
$\overline{12131312}$ where the particle bounces off disk 1 instead of
moving directly between disk 2 and disk 3.  This orbit also bifurcates at
$r_c=4.1044\ldots$ where, in the configuration space, it coincides with
the orbit of figure~\ref{f_123132}~a). 

Clearly all orbits that coincide with the orbit of 
figure~\ref{f_123132}~b)  for $r_c$ bifurcate exactly at this parameter
value, and the symbolic description of these orbits is obtained by
either including or not including the symbol for disk 1 each time the
particle is tangent to disk~1.  One example is when we chose to include
the symbol 1 each time the particle goes from disk~3 to disk 2 but not
when going from disk 2 to disk  3.  This gives the periodic orbit
$\overline{1231312}$ drawn in figure~\ref{f_123132}~d). This orbit has
the same length in configuration space as the first orbits for the
bifurcation parameter $r_c$, but it has a broken  symmetry and we can
think of it as the orbit born at a symmetry breaking bifurcation in a
smooth potential.  Examples of other periodic orbits in the bifurcation
family are  $\overline{1231321231312}$, which does not bounce at the
tangency point the first 3 times, and then bounces once, and the orbit
$\overline{123131212131312}$ that does not bounce the first time, but
then bounces the next 3 times at the tangency point.   The length of the
symbol sequence of these two orbits is approximately twice the length of
the symbol sequence of the short orbits, and for the bifurcation
parameter value $r_c$, the length of these orbits in configuration space
is exactly twice the length of the short orbits. These orbits  could be
born at period doubling bifurcations in a smooth potential.  Orbits
3,4,5,\ldots
times the length of  the original orbits are constructed the same way.

All orbits in this bifurcation family are of the form
\[   \ldots 212s_{-1}313s_{0}212s_{1}313s_{2}\ldots  \]
\noindent where $s_i$ stands for a symbol 1 or no symbol.  Periodic
sequences of $s_i$ correspond to periodic orbits while non-periodic
sequences of $s_i$ correspond to chaotic orbits.

We observe that the scenario of bifurcations is similar to the tent map;
all bifurcations take place at one parameter value $r_c$.  The
number of orbits in a family here is however larger than in the
one dimensional unimodal map period doubling scenario;  in one dimension
we repeated the symbol sequence twice and changed the last symbol, while
here we repeat a short symbol sequence and then have different
possibilities for changing a symbol.   We expect the
bifurcations in a smooth potential to be more complicated and with a
many dimensional parameter space.

To compare a billiard system with the smooth Hamiltonian (\ref{e_H}) we
investigate the bifurcation family of the orbit
$\overline{1(32)^44(23)^4}$ in the 4-disk billiard shown in
figure~\ref{f_13244234}~a).   
The parameter in the 4-disk billiard is 
$r=r_{12}\equiv r_{23}\equiv r_{34}\equiv r_{41}$.
This orbit bifurcates when the line from disk
1 to disk 3 is tangent to disk 2, and because of the symmetry of the
orbit, at the same parameter value the line from disk 4 to disk 2 is tangent to
disk 3.  The orbit that bounces off each of the tangent points is the
orbit $\overline{12(32)^4343(23)^42}$ drawn  in
figure~\ref{f_13244234}~b).  We can now construct all possible
combinations where the orbit does or does not bounce off the disk 2 or 3 at the
tangent points.  The shortest possibilities are: 
$\overline{1(32)^434(23)^42}$,  $\overline{1(32)^4343(23)^4}$,
$\overline{1(32)^44(23)^42}$, $\overline{1(32)^443(23)^42}$ and
$\overline{1(32)^4343(23)^42}$, and orbits symmetric to these. The
general form of this bifurcation family of orbits is
\begin{equation}
   \ldots
     1s_{-1}(32)^4t_{-1}4t_0(23)^4s_01s_{1}(32)^4t_{1}4t_2(23)^4s_2
   \ldots    \label{e_family}
\end{equation}
with $s_i$ either 2 or no symbol, and $t_i$ either 3 or no symbol.

The position on the edge of disk 1 for the bounce, is shown in
figure~\ref{f_pos_d_1} for some of these orbits as a function of the
parameter $r$.  The critical parameter value is $r_c = 2.0312\ldots $ and
the orbits are not physically admissible for $r<r_c$.  Figure
\ref{f_pos_d_1} shows that when $r>r_c$ the positions spreads out with
orbit $\overline{1(32)^44(23)^4}$ having the largest value and the orbit
$\overline{12(32)^4343(23)^42}$ having the smallest value.  

The bifurcation diagram for some of the orbits in this family for the
Hamiltonian  (\ref{e_H}) is described in ref.\rf{DR} and is sketched in
figure~\ref{f_bif_diagr}. Only the short orbits are included and the
drawing is not correctly scaled in the parameter value, and  the
horizontal positions are chosen only for better visualization (for numerical
values see figure 4 in ref.\rf{DR}).  A dotted line is a hyperbolic
orbit while a solid line is an elliptic orbit. The figure shows that the
two orbits $\overline{1(32)^44(23)^4}$ and 
$\overline{12(32)^4343(23)^42}$ are born together, and that 
$\overline{1(32)^44(23)^4}$ is the hyperbolic orbit while
$\overline{12(32)^4343(23)^42}$ is the elliptic orbit.  The elliptic
orbit then has bifurcations where it  bifurcates into orbits belonging
to the family (\ref{e_family}).   This family structure of symbols for
this Hamiltonian (\ref{e_H}) is conjectured by Dahlqvist and
Russberg\rf{DR}.  By finding longer orbits we expect each elliptic
branch of the tree to go through an accumulation of bifurcations similar
to the logistic map.   There is no reason to expect our choice of the
parameter $a$ to be in any way a universal choice, or that this should
be a universal one-dimension parameter bifurcation scenario like for the
unimodal map.  A different  parametrization of the Hamiltonian may give
a slightly different bifurcation tree, but  always including exactly the
orbits in the family (\ref{e_family}).     The number of orbits in the
bifurcation family is larger than the period doubling family in the
logistic map, and this implies 
that some orbits have to bifurcate several times.  For example, in  the
logistic map no orbits bifurcates more than once, while in the example
above the orbit $\overline{12(32)^4343(23)^42}$ bifurcates three times.

The example shows that if we know the bifurcation family in a billiard
system we can predict the expected bifurcations in a smooth potential
and the geometrical shape of the orbits created in a bifurcation.  We
expect a potential with 3 or 4 smooth hills to exhibit bifurcation trees
of similar structure.  The theory of bifurcation in smooth 
potentials\rf{meyer,greene_et_al,baranger} predicts bifurcations when a
eigenvalue $\lambda$ of the monodromy matrix has rational phase on the
unit circle, or when it becomes real.  The conservation of the
Poincar\'e index gives some restrictions on the possible bifurcations. 
The theory does however not predict the geometrical structure of the
bifurcating orbits, nor tells anything  about how many times an orbit
bifurcates with the same eigenvalue.  Study of the billiard bifurcations
adds this information to the theory.
This may give a unique symbolic description of all unstable and stable
orbits in a potential.

An important implication of the billiard studies is that if the
structure of bifurcations in the 4-disk  billiard and in the smooth
potential are similar, then we expect to be able to describe the allowed
orbits in the smooth potential by a pruning front similar to the pruning 
front for the billiard\rf{hansen_1,hansen_2}.
This fives a method to find an approximation of a Markov partition of
the system which is useful in calculations of thermodynamical quantities

\section{Conclusions}

We have here illustrated the utility of comparing bifurcations in
billiard systems and in smooth Hamiltonian systems by an example
comparing disk billiards and the $(x^2 y^2)^{1/a}$ potential. The
bifurcations is easy to find in billiards and we obtain the pruning
front which determines the forbidden orbits and the symbolic dynamics
for the system.  It is more difficult to find the bifurcations in a
smooth potential and the existence of a pruning front in that case is
not yet established.  We have shown here that by using symbolic 
dynamics we can find bifurcation families of orbits and that these
families have the same structure in the billiard and in the smooth
potential.  This  indicates that the pruning front is similar for the
billiards and for  smooth potentials.  Knowledge of the bifurcations in
billiards can also be of help in numerical searches for bifurcations in
a potential.

\vfill\eject

\begin{enumerate}

\fg{f_123132}
Examples of orbits that belongs to the same bifurcation family for the 3
disk billiard. 
a) The orbit $\overline{123132}$ with $r_{12}=r_{13}=r_{23}=2.5$.
b) The orbit $\overline{123132}$ with $r_{12}=r_{13}=2.5$ and
$r_{23}=r_c=4.1044\ldots$; the bifurcation point for the orbit.
c) The orbit $\overline{12131312}$, $r_{12}=r_{13}=r_{23}=2.5$.
d) The orbit $\overline{1231312}$, $r_{12}=r_{13}=r_{23}=2.5$.

\fg{f_13244234}
Examples of orbits that belong to the same bifurcation family in the 4
disk billiard with $r=2.5$. 
a) The orbit $\overline{1(32)^44(23)^4}$. 
b) The orbit $\overline{12(32)^4343(23)^42}$. 
c) The orbit $\overline{1(32)^434(23)^42}$. 
d) The orbit $\overline{1(32)^4343(23)^4}$. 

\fg{f_pos_d_1} 
The position on disk 1 of a bounce as a function of the parameter $r$
for some periodic orbits in the 4 disk billiard.  The bifurcation point
is $r_c = 2.0312\ldots$.  The positions for $r>r_c$ of the two orbits
$\overline{1(32)^44(23)^4}$ and $\overline{1(32)^4343(23)^4}$ are close
on the uppermost curve and the two orbits
$\overline{12(32)^4343(23)^42}$ and $\overline{12(32)^44(23)^42}$ are
close on the down-most curve.  For $r<r_c$ all the (physically not
admissible) orbits are close.

\fg{f_bif_diagr}
A sketch of the bifurcation diagram of the Hamiltonian (1).   Solid
lines correspond to elliptic orbits and dotted lines to hyperbolic
orbits. Only short orbits are included.

\end{enumerate}

\vfill\eject

{\bf Acknowledgements: }
The author is grateful to P. Cvitanovi\'c, P. Dahlqvist, G. Russberg and
the colleagues in  the Chaos group at the Niels Bohr   Institute for
discussions. The author thanks the Norwegian Research Council for
support.


\renewcommand{\baselinestretch} {1}

\end{document}